\documentclass[sigconf]{acmart}
\usepackage{array}
\newcolumntype{P}[1]{>{\centering\arraybackslash}p{#1}}
\usepackage{enumitem}
\usepackage{booktabs}
\usepackage{fancybox}

\AtBeginDocument{%
  \providecommand\BibTeX{{%
    \normalfont B\kern-0.5em{\scshape i\kern-0.25em b}\kern-0.8em\TeX}}}

\copyrightyear{2021}
\acmYear{2021}
\setcopyright{acmcopyright}\acmConference[EASE 2021]{Evaluation and Assessment in Software Engineering}{June 21--23, 2021}{Trondheim, Norway}
\acmBooktitle{Evaluation and Assessment in Software Engineering (EASE 2021), June 21--23, 2021, Trondheim, Norway}
\acmPrice{15.00}
\acmDOI{10.1145/3463274.3463338}
\acmISBN{978-1-4503-9053-8/21/06}

\begin{document}

\title{From Blackboard to the Office: A Look Into How Practitioners Perceive Software Testing Education}

\author{Luana Martins}
 \orcid{https://orcid.org/0000-0001-6340-7615}
 \affiliation{%
   \institution{Federal University of Bahia}
   \city{Salvador}
   \state{Bahia}
   \country{Brazil}}
    \email{martins.luana@ufba.br}
   
\author{Vinícius Brito}
 \affiliation{%
   \institution{Federal University of Bahia}
   \city{Salvador}
   \state{Bahia}
   \country{Brazil}}
 \email{viniciusbj@ufba.br}

\author{Daniela Feitosa}
 \affiliation{%
   \institution{Federal University of Bahia}
   \city{Salvador}
   \state{Bahia}
   \country{Brazil}}
 \email{dfeitosa@ufba.br}
 
\author{Larissa Rocha}
 \affiliation{%
   \institution{State University of Feira de Santana}
   \city{Feira de Santana}
   \state{Bahia}
   \country{Brazil}}
 \email{lrsoares@uefs.br}
 
\author{Heitor Costa}
 \affiliation{%
   \institution{Federal University of Lavras}
   \city{Lavras}
   \state{Minas Gerais}
   \country{Brazil}}
 \email{heitor@ufla.br}

 \author{Ivan Machado}
 \affiliation{%
   \institution{Federal University of Bahia}
   \city{Salvador}
   \state{Bahia}
   \country{Brazil}}
 \email{ivan.machado@ufba.br}

\renewcommand{\shortauthors}{Martins, et al.}
\begin{abstract}

The teaching-learning process may require specific pedagogical approaches to establish a relationship with industry practices. Recently, some studies investigated the educators' perspectives and the undergraduate courses curriculum to identify potential weaknesses and solutions for the software testing teaching process. However, it is still unclear how the practitioners evaluate the acquisition of knowledge about software testing in undergraduate courses. This study carried out an expert survey with 68 newly graduated practitioners to determine what the industry expects from them and what they learned in academia. The yielded results indicated that those practitioners learned at a similar rate as others with a long industry experience. Also, they studied less than half of the 35 software testing topics collected in the survey and took industry-backed extracurricular courses to complement their learning. Additionally, our findings point out a set of implications for future research, as the respondents' learning difficulties (e.g., lack of learning sources) and the gap between academic education and industry expectations (e.g., certifications).

\end{abstract}



\maketitle

\section{Introduction}
\label{sec:Introduction}

The industry-academia collaboration lack is widely discussed in the literature \cite{Wohlin2013, Petersen2014, Kanso2014, 10.1145/2647648.2647656, 10.1145/3195546.3195552, 10.1145/2647648.2647651}. Besides limiting the academic research practical application, that lack may impact the teaching-learning process. Existing teaching methods may not be enough to support practitioners in addressing practical software engineering challenges \cite{Arcuri2018}. It is a long-term and widespread issue in the software engineering field. Newly graduated practitioners face several challenges to apply academic knowledge to a practical context.  For example, they should communicate effectively with their teams to report problems using specific tools and techniques. However, the academic teaching-learning process lacks practical classes that could improve the necessary skills to perform software testing activities, like communication and teamwork \cite{Melo2020, scatalon2018}.

In particular, we have observed shreds of evidence supporting this claim regarding the software testing area \cite{yamashita2015, Garousi2016, GarousiHerkiloglu2016, GarousiFelderer2017}. Software companies have faced difficulties selecting qualified personnel for performing testing activities \cite{Astigarraga2010, Melo2020}. Such a scenario is likely due to the way academia teaches software testing \cite{Valle2015}. Several studies investigate different software testing perspectives to fulfill the industry-academia gap. \citet{GarousiFelderer2017} proposed selecting relevant topics for the industry-academia collaboration based on the practitioners’ perspective. \citet{Melo2020} investigated the educators’ perspective regarding the syllabus, materials, practices, and challenges of the software testing teaching. \citet{scatalon2018} analyzed the graduates’ curriculum to determine the lack of content extracted from a textbook and the practitioners’ activities. However, they do not focus on the practitioners’ perception of their software testing learning in academia.

Our study aims to support the software testing community by discussing how Brazilian practitioners perceive software testing education. We conducted an expert survey with 68 practitioners to investigate the software testing learning to determine whether there is a clear difference between (i) what the industry expects from newly graduated practitioners, from now on called \textit{novice software testers}, and (ii) what they learn while undergraduate students. The survey addresses the following research questions:

\begin{enumerate}[label=\bf RQ\arabic*:,leftmargin=.8cm]
    \item \textbf{How do testers assess their academic learning in software testing?} This RQ investigates the test practitioners’ view of what they learned in academia to analyze whether the content was relevant to develop your career.
    
    \item \textbf{~Which theoretical and practical aspects of the software testing do testers perceive as relevant to teach in academia?} This question investigates the test practitioners’ view of what they use in their daily testing activities that might be interesting to teach in academia.
    
    \item \textbf{What are the industry’s expectations concerning novice software testers?} This question investigates the practitioners’ perception about what skills the industry expects those testers have to start working as software testing professionals.
    
    \item \textbf{What is the novice software testers’ knowledge of software testing?} This question investigates the test practitioners’ view of how academia trains testers.
\end{enumerate}

Our study's main findings are: (i) the learning rate of novice software testers is similar to the learning rate of experienced professionals; (ii) software testers studied in academia less than a half of the contents they consider important in practice; (iii) novice software testers usually take industry-backed extracurricular courses to complement their learning. Besides, our study provides a list of the main challenges the practitioners face to learn about software testing in academia and their challenges for the industry entrance. Although our study presents the Brazilian scenario, other countries may observe a similar scenario.
\section{Background}
\label{sec:Background}

Software companies face the challenge of developing complex, high-quality systems without increasing project development costs. In this context, software testing is essential to ensure software quality \cite{spadini2018, tassio2019}. However, the lack of qualified human resources is a barrier to performing software testing activities in software companies \cite{Astigarraga2010, Valle2015, Melo2020}. That fact might be related to the difficulty and inefficiency in the software testing teaching-learning process \cite{GarousiFelderer2017, GarousiHerkiloglu2016, Garousi2016, yamashita2015}. 
Software testing is a practical discipline that faces challenges to relate concepts and practices between the industry needs and the contents approached in academia. The graduated professionals may not acquire the needed technical (e.g., programming knowledge) and personal skills (e.g., communication and teamwork) \cite{Radermacher2013, scatalon2018}, which can hinder the performance of the software testing activity. Therefore, the software testing teaching cannot be enough to bring future professionals closer to the industry practices \cite{Lethbridge2007, Melo2020, scatalon2018}. Students may face difficulties in applying software testing practices, as academia does not teach them how to communicate effectively for problem-solving through real examples, using different tools and techniques. 

Several studies have investigated the universities' contribution to test practitioners' preparation \cite{Melo2020, scatalon2018} to overcome the problems generated in the teaching-learning process. From a global perspective, there are differences in software testing teaching. Software engineering courses include software testing contents in either elective or mandatory disciplines. Most software testing disciplines have a workload ranging from 6 to 88 hours, focusing on teaching different software testing levels, types, and techniques~\cite{Melo2020}. 

In the Brazilian context, software engineering courses encompass the software testing content based on the Association for Computing Machinery (ACM) curriculum guidelines~\cite{ACM2015}. However, the courses usually dedicate less than ten hours to teach software testing, without any perspectives of integrating it to other disciplines \cite{Valle2015}. The Brazilian curricular guidelines indicate software testing as a curricular component desirable to teach, but the universities have left the decision to select the contents to teach \cite{benitti2012}. \citet{benitti2012} analyzed the curricula from top-rated universities in Brazil and identified that software testing contents commonly encompass the following categories:

\begin{itemize}

    \item \textbf{Testing Fundamentals}. This category comprises the testing terminology and concepts;
    
    \item \textbf{Testing levels}. This category comprises unit, integration, system and acceptance testing, and alpha-beta testing;
    
    \item \textbf{Testing techniques}. This category comprises white-box and black-box testing, model-based testing, code-based testing, technical revision, static analysis, structural testing, mutation testing, and test coverage;
    
    \item \textbf{Testing types}. This category comprises human interface testing, web application testing, and regression testing;
    
    \item \textbf{Testing process}. This category describes the testing process using documentation, software quality assurance, and formalized processes.
    
\end{itemize}
\section{Survey}
\label{sec:ResearchMethodology}

In this study, we employed survey research using a sample of 68 software testing practitioners from Brazil. This section introduces the target audience identification and characterization, the survey design, the participant selection criteria, pilot test, data collection, and qualitative data analysis. The study design followed the guidelines defined by \citet{kitchenham2002}.

\subsection{Target Audience}

The practitioners needed to relate the academic concepts of software testing with the industry needs. Therefore, our sample's composition only included practitioners in software testing who hold a Bachelor's degree in a computer science-related course.

\subsection{Questionnaire Design}

The first questionnaire statement was: \textit{``I have experience in software testing and graduated in a computing-related course''.} If the respondents answered \textit{``NO''} for the question, they did not continue answering the questionnaire, and the questionnaire ended. Otherwise, we presented 20 questions grouped into five categories (\textbf{Table~\ref{table:questionnaire}}):

\begin{enumerate} 

    \item \textbf{Respondents' profile and demographics.}  This category contains five questions about the respondents' undergraduate course and experience in software testing; 
    
    \item \textbf{Respondents' perception about academic learning in software testing}. This category investigates the RQ1. Before answering the seven mandatory questions from this category, the respondents inform whether they learned about software testing in academia. The question is, \textit{"Did you learn about software testing in any undergrad course?"}. If the answer was \textit{"YES"}, they had to answer all the mandatory questions on their perception of what they learned in academia. Otherwise, the questionnaire led them to the next category; 
    
    \item \textbf{Respondents' perception about needed software testing theoretical and practical aspects for teaching in academia}. This category investigates the RQ2 and contains three mandatory questions regarding the software testing relevant types, approaches, and techniques to teach in academia. Also, the respondents should indicate tools and suggestions to improve the academic teaching in software testing;
    
    \item \textbf{Respondents' perception about the testers' profile: expectation vs. reality}. This category investigates RQ3 and RQ4 and contains one optional and three mandatory questions. They investigate the gap between the testers' profile that the industry expects from novice software testers and the software testing content that the academia educates;
    
    \item \textbf{Acknowledgments}. This category presents a thankful message for the respondents' participation. Additionally, it contains one question to know whether we could contact them for further interaction.
    
\end{enumerate}

\begin{table}[t]
\footnotesize
\centering
\caption{Survey questionnaire}
\label{table:questionnaire}
\vspace{-4pt}
\begin{tabular}{P{0.3cm}p{6.0cm}P{0.7cm}}
\toprule
\multicolumn{1}{c}{\#} & \multicolumn{1}{c}{Question} & \multicolumn{1}{c}{Answer} \\ \midrule
1 & How long  did  you  finish  your  graduation  course  in  computing-relate  area? & C \\
2 & In which state did you finish your graduation course in computing-related area? & C \\
3 & During  your  graduation  course,  you  attended  a public or private university? & C \\
4 & Which  course  did  you  graduate  in  computing-related  area? & C \\
5 & How many years of experience  do  you  have  in  the  software  testing  area? & C \\
6 & Did  you  learn  about  software  testing  in  any  discipline  in  your  graduation  course? & C \\
7 & In  your  opinion,  how  relevant  was  the  test  content  taught  in  the  university  to  the development  of  your  profession? & L \\
8 & Please  justify  the  grade  given  in  the  previous  question. & O \\
9 & Regarding  the  content  covered  on  software  testing,  do  you  consider  that  it  was more  theoretical  or  more  practical? & L \\
10 & What  topics  related  to  software  testing  do  you  remember  studying  in  your graduation  course? & O \\
11 & What  difficulties  did  you  face  to  learn  the  software  testing  topics? & O \\
12 & How much do you apply the knowledge  acquired  in  the  university  for  software testing? & L \\
13 & What  difficulties  did  you  face  to  apply  the  knowledge  acquired  in  the  university to  the  practice  of  software  testing  in  the  industry? & O \\
14 & What types, approaches and testing techniques do you think should be addressed at the university? & L \\
15 & What  types,  approaches  and  testing  techniques  not  previously  listed  do  you think  that  should  be  addressed  in  academia? & O \\
16 & What  tools  and  methodologies  do  you  think  are  useful  for  developing  tests  in the  industry? & O \\
17 & What  do  you suggest  to  improve  the  teaching  of  software  testing? & O \\
18 & In  your  opinion,  considering  the  testing  activity,  what  does  the  industry  expect from  a  newly  graduate? & O \\
19 & Did  you  have  specific  training  in  software  testing  to  perform  the  activities assigned  to  you?  If  so,  could  you  describe  it? & O \\
20 & In  your  opinion,  considering  the  software  testing  activity,  are  newly  graduates prepared  to  work  in  the  industry?  If  not,  could  you  explain? & O \\ \bottomrule 
\end{tabular} 
\caption*{{\footnotesize{C: Closed-question, O: Open-question, L: Likert-scale}}}
\vspace{-16pt}
\end{table}

\subsection{Pilot Test}

We tested the questionnaire with respondents from the target audience to improve its understandability and remove any inconsistencies. The pilot questionnaire was applied on November 10th, 2020, and counted on five respondents. It comprised questions to evaluate the significance and respondents' interest in the research objectives. Also, we collect data about the time to complete the questionnaire and the respondents' feelings and satisfaction about the research process. As a result, the respondents considered the number and content of questions adequate to the research objectives and spent 15 minutes on average in this task. They identified some minor issues: need to number the questions, change the answers' types (e.g., from multiple-choice to check-box), improve the readability of some questions, and add one additional question to inform where they learned about software testing.

\subsection{Distribute the Questionnaire}

We sent invitations by e-mail and professional networking social media (\textit{e.g.}, LinkedIn) for the Brazilian practitioners whose profile meets the target audience selection criteria. Thus, our method was \textit{convenience sampling}, a dominant experimental approach in the software engineering field \cite{Sjoeberg2005, GarousiFelderer2017}. We sent the invitations from January 1st to January 19th, 2021. The invitation provided necessary information about the study purpose, choice justification, and the respondents' participation importance. Besides, we informed potential participants of the study privacy policies in a clear and detailed manner. We sent invitations to 145 practitioners and received 68 answers (response rate of 46\%).

\subsection{Results Analysis and Report Writing}

We adopted four assumptions in the questionnaire to achieve the defined objectives:

\begin{enumerate}
    
    \item We applied single closed-questions with a range of values to characterize the target audience. When discussing the results, the sum of percentages is equal to a hundred percent. 
    
    \item We applied a five-point Likert Scale from ``Irrelevant'' (1) to ``Very Important'' (5) to ask about the practitioners' perception of software testing. When discussing the results, the sum of percentages is equal to a hundred percent. 
    
    \item We conducted the open coding process for the open questions. For example, the respondent \#R7 said that the industry expects from a newly graduated ``willingness to learn, perception of details and creation of good test cases''. Three authors independently extracted the general concepts from the answers and discussed the divergences.  For that case, they identified two codes: soft skills and test case development. 
    
    \item We selected excerpts from the open questions answers to support the discussion of the results. We used a unique identifier for respondents' each of the quotes. For example, \#R1 indicates the first respondent's answers.
    
\end{enumerate}
\section{Results}
\label{sec:Results}

In this section, we reported the survey results, which consists of the respondents’ profile and demographics, academic learning assessment (RQ1), relevant topics to the teaching in the academia (RQ2), the novice software testers’ profile (RQ3), and novice software testers’ profile expected by the industry (RQ4). Data is publicly available in an online open data repository \cite{data}.

\begin{figure*}[t]
\centerline{\includegraphics[width=0.9\linewidth]{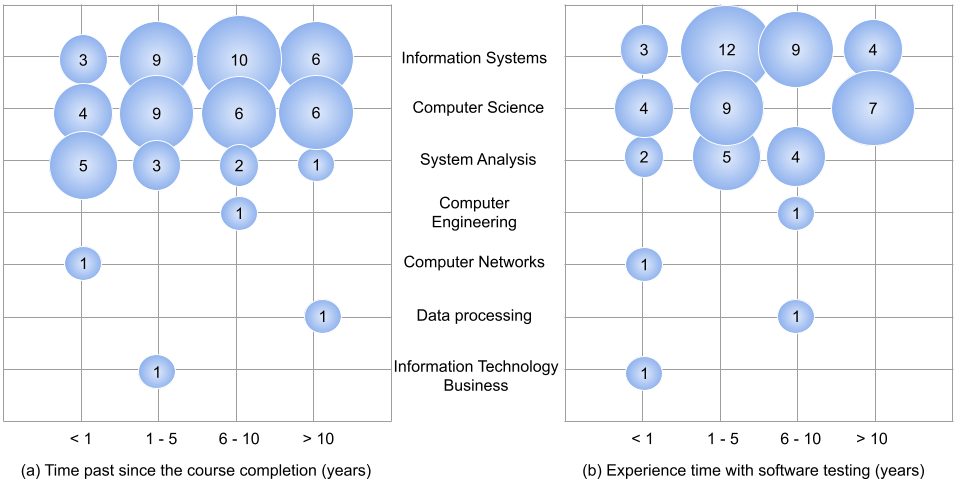}}
\caption{Distribution of the respondents by experience and course completion}
\label{fig:graduation_time}
\end{figure*}

\subsection{Respondents' Profile and Demographics }

The respondents graduated from Brazilian universities, where 26 respondents studied in public universities (38.2\%), and 42 respondents studied in private universities (61.8\%). Besides one university from Federal District, the respondents graduated from twelve different states (out of twenty-six states), representing the five Brazilian regions. The region with the lesser amount of respondents was the Central West (only one respondent - 1.5\%), and the region with the higher number of respondents was the Southeast (42 respondents - 61.7\%). Those results are similar to what we expected due to the distribution of public and private universities in Brazil. According to the last census, \cite{nunes2016educaccao}, the five Brazilian regions have computing-related courses following the distribution: Southeast region - 47.3\%, South region - 21.9\%, Northeast region - 16.6\%, North region - 4.4\%, and Central West region - 9.8\%.  

Additionally, we asked the respondents about their undergraduate courses. We obtained the following responses: 28 respondents had a degree in Information Systems (41.1\%); 25 respondents had a degree in Computer Science (36.7\%); 11 respondents in System Analysis  (16.2\%); and one respondent in each of the following courses, Computer Engineering (1.5\%), Information Technology Business (1.5\%), Data Processing (1.5\%), and Computer Networks (1.5\%).
Regarding the time they hold a degree,
13 respondents completed their courses less than one year ago (19.1\%), 22 respondents completed it between one and five years ago (32.4\%), 19 respondents completed it between five and ten years ago (27.9\%), and 14 respondents completed the course more than ten years ago (20.6\%). \textbf{Figure \ref{fig:graduation_time}(a)} shows respondents’ distribution by undergraduate course and the period they hold the degree.

Regarding the respondents’ experience with software testing, 11 respondents work less than one year (16.2\%), 26 respondents work between one and five years (38.2\%), 20 respondents work between five and ten years (29.4\%), and 11 respondents worked more than ten years (16.2\%). \textbf{Figure \ref{fig:graduation_time}(b)} shows the respondents’ experience by their undergraduate course. When comparing the respondents’ experience and time of course completion, especially for the Computer Science and System Analysis courses, some respondents started working during their undergraduate course. Therefore, the number of respondents in an experience range can be higher than the time of course completion in the same range.

\vspace{-6pt}
\subsection{Academic Learning Assessment (RQ1)}

We asked respondents about their perception of the software testing teaching in their undergraduate courses. Since only 38 respondents studied software testing (55.9\%),  we considered only those respondents’ answers in this section.

We observed the university type (public or private), computing-related course, and how long they hold a degree to understand why only half of the respondents studied software testing in academia. The number of respondents in public or private universities decreased at a similar rate, 38 from 68 studied testing topics: 15 respondents (in a total of 26) from public universities (39.5\%) and 23 respondents (in a total of 42) from private universities (60.5\%). The number of respondents also proportionally decreased more than 50\% for all course completion ranges: 7 respondents (in a total of 13) with less than one year (53.9\%); 13 respondents (in a total of 22) from 1 to five years (59.1\%); 10 respondents (in a total of 19) from 6 to 10 years (52.6\%), and eight respondents (in a total of 14) with more than ten years (57.1\%) of course completion. 

\vspace{0.2cm}
\doublebox{
\begin{minipage}[c]{23em}
\textit{\textbf{Finding 1.}} Only 55.9\% of respondents studied software testing in academia. We have no evidence to suggest that it occurs due to the university type (public or private), or time of course completion.
\end{minipage}
}
\vspace{0.2cm}

\vspace{0.2cm}
\doublebox{
\begin{minipage}[c]{23em}
\textit{\textbf{Finding 2.}}
About half of the respondents indicated they did not study software testing while they were undergraduate students. It may indicate that some computing-related courses do not teach software testing at all.
\end{minipage}} 
\vspace{0.5cm}

Additionally, we asked the respondents to remember the studied software testing topics in their undergraduate courses (\textbf{Table \ref{table:topics_remembered}}). They mentioned 24 topics, where each respondent could cite one or more topics. Also, we asked about the difficulties they faced to learn those topics (one optional question), where six respondents did not answer (15.8\%), and 32 respondents answered one or more difficulties (84.2\%) (\textbf{Table \ref{table:Leaning_Difficulties}}). In general, they mentioned ten difficulties, but most of them expressed the same difficulty:  applying the concepts they learned in theory (13 respondents - 40.6\%).

\vspace{0.6cm}
\doublebox{
\begin{minipage}[c]{23em}
\textit{\textbf{Finding 3.}}
Unit Testing, Software Testing Concepts, White-box Testing, Black-box Testing, and Test Types were the most remembered topics.
\end{minipage}} 

\begin{table}[b]
\centering
\footnotesize
\caption{Studied topics}
\vspace{-4pt}
\label{table:topics_remembered}
\begin{tabular}{P{0.3cm}p{4.2cm}P{1.3cm}P{1.3cm}}
\toprule
\# & \multicolumn{1}{c}{Topics on Software Testing} & \#Answers & Percentage \\
\midrule
1 & Unit Testing & 12 & 31.6\% \\
2 & Black-box testing & 9 & 23.7\% \\
3 & Software Testing Concepts & 9 & 23.7\% \\
4 & Test Types & 9 & 23.7\% \\
5 & White-box Testing & 9 & 23.7\% \\
6 & Functional Testing & 6 & 15.8\% \\
7 & Integration Testing & 4 & 10.5\% \\
8 & Software Life-cycle & 4 & 10.5\% \\
9 & Tools & 4 & 10.5\% \\
10 & Automation Testing & 3 & 7.9\% \\
11 & Test Cases & 3 & 7.9\% \\
12 & Coverage & 2 & 5.3\% \\
13 & Quality process & 2 & 5.3\% \\
14 & Tests Importance & 2 & 5.3\% \\
15 & Usability Testing & 2 & 5.3\% \\
16 & Acceptance Testing & 1 & 2.6\% \\
17 & Non-functional Testing & 1 & 2.6\% \\
18 & Performance Testing & 1 & 2.6\% \\
19 & Regression Testing & 1 & 2.6\% \\
20 & Security Testing & 1 & 2.6\% \\
21 & System Testing & 1 & 2.6\% \\
22 & Test Planning, Management and Maintenance & 1 & 2.6\% \\
23 & Test Strategies & 1 & 2.6\% \\
24 & Verification and Validation & 1 & 2.6\% \\
\bottomrule
\end{tabular}
\end{table}

\begin{table}[b]
\footnotesize
\caption{Difficulties in learning}
\vspace{-4pt}
\label{table:Leaning_Difficulties}
\begin{tabular}{P{0.2cm}p{4.2cm}P{1.2cm}P{1.2cm}}
\toprule
\# & \multicolumn{1}{c}{Difficulties} & \multicolumn{1}{c}{\#Answers} & \multicolumn{1}{c}{Percentage} \\
\midrule
1 & Concepts application & 13 & 40.6\% \\
2 & Learning sources & 9 & 28.1\% \\
3 & Concepts understanding & 4 & 12.5\% \\
4 & Tools usage & 3 & 9.4\% \\
5 & Dedication & 1 & 3.1\% \\
6 & Didactic & 1 & 3.1\% \\
7 & Functional & 1 & 3.1\% \\
8 & Programming & 1 & 3.1\% \\
9 & Testing Design & 1 & 3.1\% \\
10 & Testing Process & 1 & 3.1\% \\
\bottomrule
\end{tabular}
\end{table}

\begin{figure}[t]
\centerline{\includegraphics[width=1.0\linewidth]{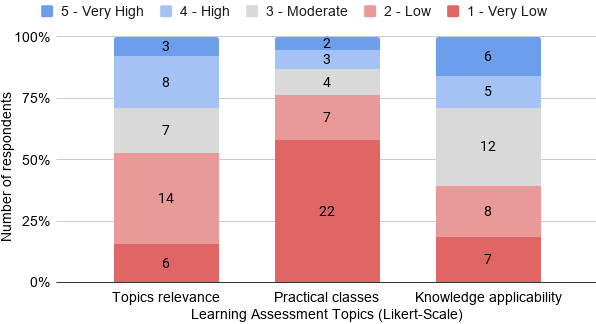}}
\caption{Relevance for career}
\label{fig:learning_assessment}
\end{figure}

\vspace{0.2cm}
\doublebox{
\begin{minipage}[c]{23em}
\textit{\textbf{Finding 4.}}
The respondents' most significant difficulties were related to applying concepts in practice and finding learning sources to support their learning process.
\end{minipage}} 
\vspace{0.4cm}

The respondents evaluated their software testing learning regarding the testing topics' relevance for their career. 
\textbf{Figure \ref{fig:learning_assessment}}
shows that more than half of the respondents considered the topics as low relevant. The 20 respondents who rated either 1 (very low) or 2 (low relevance) listed some negative aspects. Fifteen out of them answered that the contents are superficial (39.5\%). Six respondents only studied theoretical contents (15.8\%). Three respondents learned software testing within other software engineering topics (7.9\%), and one respondent considered the topics as not applicable in practice (2.6\%). 

One respondent had experience in software testing before the undergraduate course (2.6\%). Nine respondents rated from 3 to 5 (moderate to very high relevance) and listed negative aspects, as follows: six respondents mentioned that the contents were superficial (15.8\%), four respondents said theoretical contents (10.5\%), and one respondent noted that the contents were not applicable in practice (2.6\%). 

Besides the negative aspects, five respondents rated as 4 or 5 points (high to very high relevance) listed some positive aspects. Seven respondents highlighted that the contents were fundamental to support their learning (18.4\%). Two respondents mentioned that they learned the topics in non-mandatory classes (5.3\%). One respondent had practical classes, experienced professors from the industry, and noted that the testing contents guided their software testing career (2.6\% each). The respondents could provide more than one justification or even abstain.

\vspace{0.4cm}
\vspace{0.2cm}
\doublebox{
\begin{minipage}[c]{23em}
\textit{\textbf{Finding 5.}}
Academia teaches few topics applicable to the software testing career. The main reasons are the superficial and theoretical contents.
\end{minipage}} 
\vspace{0.4cm}

Subsequently, the respondents rated the software engineering content as either theoretical or practical and how much the acquired knowledge was applicable when executing testing tasks. \textbf{Figure \ref{fig:learning_assessment}} shows that more than 75\% of respondents had theoretical classes (theory), and 40\% of respondents considered the topics of low applicability in practice. Additionally, we questioned them about the difficulties of applying the concepts learned in academia in the industry (\textbf{Table \ref{table:Applying_Difficulties}}). We noticed the company standards refer to the time dedicated to the software testing tasks, i.e., besides knowing the testing process, the company standards limited the respondents.

\begin{table}[t]
\footnotesize
\caption{Difficulties in applying concepts}
\label{table:Applying_Difficulties}
\begin{tabular}{P{0.2cm}p{4.2cm}P{1.2cm}P{1.2cm}}
\toprule
\# & \multicolumn{1}{c}{Difficulties} & \multicolumn{1}{c}{\#Answers} & \multicolumn{1}{c}{Percentage} \\
\midrule
1 & Lack of practical experience & 12 & 31.6\% \\
2 & Insufficient content & 7 & 18.4\% \\
3 & Company standards & 6 & 15.8\% \\
4 & Difficulties in learning & 3 & 7.9\% \\
5 & Tools usage & 1 & 2.6\% \\
6 & Learning sources & 1 & 2.6\% \\
\bottomrule
\end{tabular}
\end{table}

\vspace{0.4cm}
\doublebox{
\begin{minipage}[c]{23em}
\textit{\textbf{Finding 6.}}
The software testing-related topics the academia addresses presented low applicability in practice. The main reasons were the lack of practice at the academia and insufficient content on software testing.
\end{minipage}} 
\vspace{0.4cm}

\subsection{Topics to teach in the academia (RQ2)}

\begin{figure*}[t]
\centerline{\includegraphics[width=1.0\linewidth]{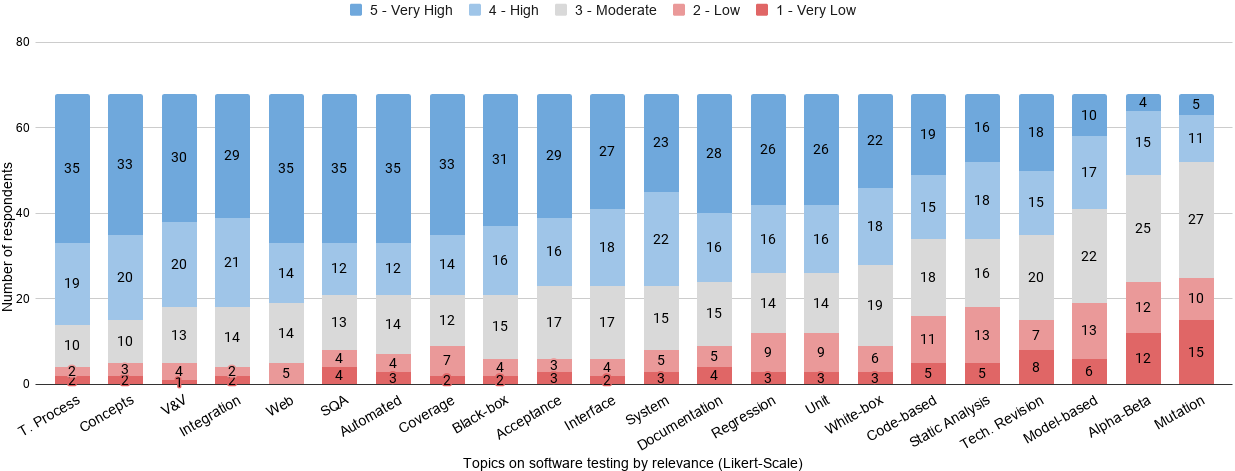}}
\caption{Software testing rating by relevance}
\label{fig:rating_topics}
\end{figure*}

We asked the respondents which topics they wish they had learned in academia (\textbf{Figure \ref{fig:rating_topics}}). More than 70\% of the respondents classified five topics as important (4 or 5 points in the Likert scale): 54 respondents indicated \textit{Testing Process} (79.4\%), 53 respondents indicated \textit{Software Testing Concepts} in general, which might be related to any other software testing topic (77.9\%), 50 respondents indicated \textit{Verification and Validation} (V\&V) and \textit{Integration Testing} (73.5\%), and 49 respondents indicated \textit{Web Testing} (72.1\%). Furthermore, more than 50\% of respondents rated other topics as important, except for 33 who indicated \textit{Technical Revision} (48.5\%), 27 indicated \textit{Model-based Testing} (39.7\%), 19 indicated \textit{Alpha-Beta Testing} (27.9\%), and 16 indicated \textit{Mutation Testing} (23.53\%). We noted that the topics considered less important received more neutral answers (3 points in the Likert scale) than negative answers (1 or 2 points in the Likert scale). Therefore, the respondents did not discard the importance of the presented topics. One additional question allowed the respondents to complement the preliminary list of the provided topics. They pointed 13 new topics, which they considered necessary (\textbf{Table \ref{table:Suggested_Topics}}). Only 29 respondents (42.6\%) answered that question. Those new topics might be related to the respondents’ daily activities.

\vspace{0.4cm}
\doublebox{
\begin{minipage}[c]{23em}
\textit{\textbf{Finding 7.}}
The respondents recognized the importance of 35 topics on software testing to teach in academia.
\end{minipage}} 
\vspace{0.4cm}

\begin{table}[t]
\footnotesize
\caption{Respondents' suggested topics}
\label{table:Suggested_Topics}
\begin{tabular}{cp{4.2cm}cc}
\toprule
\# & \multicolumn{1}{c}{Topics} & \multicolumn{1}{c}{\#Answers} & \multicolumn{1}{c}{Percentage} \\
\midrule
1 & Performance Testing & 6 & 20\% \\
2& Test-Driven Development & 4 & 13.3\% \\
3 & Stress Testing & 4 & 13.3\% \\
4 & Security Testing & 3 & 10\% \\
5 & API Testing & 2 & 6.7\% \\
6 & Mobile Testing & 2 & 6.7\% \\
7 & Usability Testing & 2 & 6.7\% \\
8 & Behavior-Driven Development & 2 & 6.7\% \\
9 & Testing Framework & 1 & 3.3\% \\
10 & Exploratory Testing & 1 & 3.3\% \\
11& Agile Methodology & 1 & 3.3\% \\
12 & Smoke Testing & 1 & 3.3\% \\
13 & Software Inspection & 1 & 3.3\% \\
\bottomrule
\end{tabular}
\end{table}

\begin{table}[t]
\caption{Learning process improvements suggestions}
\label{table:Suggestions_Improvements}
\footnotesize
\begin{tabular}{P{0.2cm}p{4.2cm}P{1.2cm}P{1.2cm}}
\toprule
\# & \multicolumn{1}{c}{Suggestions} & \multicolumn{1}{c}{\#Answers} & \multicolumn{1}{c}{Percentage} \\
\midrule
1 & Practical Classes & 27 & 39.7\% \\
2 & Specific Disciplines & 14 & 20.6\% \\
3 & Alignment with Industry Practice & 8 & 11.8\% \\
4 & Fundamental Concepts & 4 & 5.9\% \\
5 & More Content & 6 & 8.8\% \\
6 & Integration with programming Disciplines & 3 & 4.4\% \\
7 & Documentation & 2 & 2.9\% \\
8 & Experienced Professors & 2 & 2.9\% \\
9 & Real Examples & 2 & 2.9\% \\
10 & Theoretical and Practice relationship & 2 & 2.9\% \\
11 & Student's Dedication & 1 & 1.5\% \\
\bottomrule
\end{tabular}
\end{table}

\begin{figure}[t]
\footnotesize
\centerline{\includegraphics[width=1.0\linewidth]{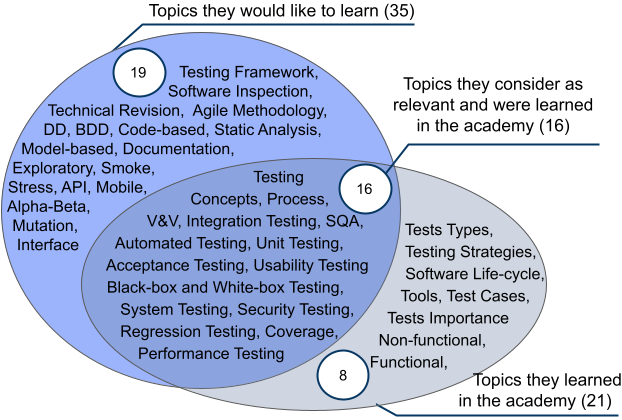}}
\caption{Comparison between the topics approached in the academia and topics collected through the survey}
\label{fig:venn_topics}
\vspace{+10pt}
\end{figure}

When comparing those results with the previous section, we identified a gap between the respondents' topics and what they remembered to have learned in academia (\textbf{Figure \ref{fig:venn_topics}}). To create the survey, we collected 22 testing topics from the literature \cite{Valle2015}. Besides them, the respondents suggested other 21 topics in the survey (43 in total). From the 43 topics, the respondents mentioned they would like to learn 35 of them in academia. However, from the 35 topics that the respondents considered important to performing testing activities in industry, they only learned 16 in academia, as the intersection in \textbf{Figure \ref{fig:venn_topics}} shows.

Additionally, we collected eight topics that the respondents mentioned they learned in academia but did not consider as important to perform software testing activities in the industry.  In \textbf{Table \ref{table:Suggestions_Improvements}}, we presented the respondents' answers when we asked what they suggest to improve the learning process. Most of the respondents mentioned they would like to have more practical classes and specific disciplines to teach software testing aligned with industry practices. For example, respondent \#R19 suggested, \textit{``It could have an up-to-date specific software testing discipline aligned with the market requirements''}.   

\vspace{0.4cm}
\doublebox{
\begin{minipage}[c]{23em}
\textit{\textbf{Finding 8.}}
Among the 35 software testing topics the respondents would like to have learned before they graduate, the results showed less than half of them in academia (45.7\%).
\end{minipage}} 
\vspace{0.4cm}

We also asked the respondents about the tools, methodologies, approaches, and techniques they consider useful for developing tests. They listed eight tools, 35 specific tools, and ten methodologies, approaches, and techniques. \textbf{Table \ref{table:Tests_Developing}} shows the respondents’ suggestions with three or more answers. According to the answers, we noticed they commonly use different tools for each organization project. One respondent did not want to suggest a tool: \textit{``I would not suggest specific tools because they are constantly evolving. However, I would use them in the teaching process as instruments to increase students' awareness of the benefits, problems, and test types''}. In this context, TestLink and Jira were the tools cited to support test case planning, documentation, and management. Teaching could benefit from such tools.

\vspace{0.4cm}
\vspace{0.2cm}
\doublebox{
\begin{minipage}[c]{23em}
\textit{\textbf{Finding 9.}}
The context of the project influences the tool selection. However, academia could adopt widely used tools to support the different learning process activities in software testing.
\end{minipage}} 
\vspace{0.4cm}

\subsection{Industry Expectations (RQ3)}

We asked respondents' opinions on the necessary skills to start working as a tester. We wanted to identify the profile that the industry expects from novice software testers. \textbf{Table \ref{table:Industry_Expectations}} presents the twelve skills listed by the respondents. As each respondent could list more than one skill, the sum of the percentages is superior to 100\%. Thus, most of the respondents mentioned \textbf{Fundamental Knowledge} and \textbf{Practical Experience}, with 30 and 38 answers, respectively (44.1\% and 41.2\%). The affirmation of respondent \#R55 supports the importance of those skills: \textit{``I believe that the industry expects a professional who can plan, analyze and execute the main types of tests [...]. The industry also expects that the professional knows the latest technologies [..]''}. Besides the skills learned through study and experience, six respondents mentioned soft skills (8.8\%). Respondent \#R32 mentioned: \textit{``The industry expects an attentive person, who can assimilate information well and have good logic.''} Therefore, it is also important that the tester be attentive, creative, proactive, dedicated, and willing to learn.

\vspace{0.4cm}
\doublebox{
\begin{minipage}[c]{23em}
\textit{\textbf{Finding 10.}}
Industry requires specific software testing knowledge, but it also expects a novice software tester to have fundamental knowledge, experience, and soft skills.
\end{minipage}} 

\begin{table}[t]
\footnotesize
\caption{Tools, methodologies, approaches, techniques useful for test development}
\label{table:Tests_Developing}
\begin{tabular}{P{0.5cm}p{3.5cm}P{1.5cm}P{1.5cm}}
\toprule
\# & \multicolumn{1}{c}{Suggestions} & \multicolumn{1}{c}{\#Answers} & \multicolumn{1}{c}{Percentage} \\
\midrule
\hline
\multicolumn{4}{c}{Types of Tools} \\
\hline
1 & Automation Testing Tools & 17 & 25\% \\
2 & Regression Testing Tools & 3 & 4.4\% \\
\hline
\multicolumn{4}{c}{Specific Tools} \\
\hline
1 & Selenium & 18 & 26.5\% \\
2 & JMeter & 8 & 11.8\% \\
3 & Jira and JUnit & 6 & 8.8\% \\
4 & Cucumber & 5 & 7.4\% \\
5 & Cypress, Postman, and TestLink & 4 & 5.9\% \\
6 & Appium and Robot framework & 3 & 4.4\% \\
\hline
\multicolumn{4}{c}{Methodologies, Approaches, and Techniques} \\
\hline
1 & Behaviour-Driven Development & 7 & 10.3\% \\
2 & Test-Driven Development & 5 & 7.48\% \\
3 & Test Pyramid Method & 3 & 4.4\% \\
\bottomrule
\end{tabular}
\end{table}

\begin{table}[t]
\footnotesize
\caption{Skills necessary to start working as a tester}
\label{table:Industry_Expectations}
\begin{tabular}{P{0.3cm}p{4.2cm}P{1cm}P{1cm}}
\toprule
\multicolumn{1}{c}{\#} & \multicolumn{1}{c}{Skills} & \multicolumn{1}{c}{Answers} & \multicolumn{1}{c}{Percentage} \\
\midrule
1 & Fundamental Knowledge & 30 & 44.1\% \\
2 & Practical Experience & 28 & 41.2\% \\
3 & Soft Skills & 6 & 8.8\% \\
4 & Programming knowledge & 7 & 10.3\% \\
5 & Test Automation & 7 & 10.3\% \\
6 & Test Case Development & 5 & 7.4\% \\
7 & Testing Management & 2 & 2.9\% \\
8 & Testing Planning and Execution & 2 & 2.9\% \\
9 & Testing Documentation & 2 & 2.9\% \\
10 & Tools & 2 & 2.9\% \\
11 & Qualified Professional & 1 & 1.5\% \\
\bottomrule
\end{tabular}
\vspace{-8pt}
\end{table}

\vspace{-10pt}
Additionally, we asked the respondents whether they needed extracurricular software testing courses to complement their academic learning. As a result, 23 respondents did not take any extracurricular course (33.8\%), and 48 respondents took at least one extracurricular course (66.2\%). Among the respondents who did an extracurricular course, nine respondents did it independently (13.2\%), and 36 respondents had training on software testing supported by the company (52.9\%). The majority of the respondents had industry-backed training, where their companies usually indicate experienced practitioners to guide newcomers. Respondent \#R61 mentioned how that scenario generally occurs:\textit{ ``I had training in pairs and in teams, where a professional taught his partner, showing in practice how to test the software''}. Conversely, companies also provide their employees with learning materials, as respondents \#R38 and \#R48 mentioned, respectively: \textit{``The company suggested courses for improvement on virtual platforms''} and \textit{``I learned from BECA, a support program for recent graduates''}. Among those respondents, only seven had certifications from associations like the International Software Testing Qualifications Board (19.4\%); and three respondents took MBA courses on software testing (8.3\%).

Furthermore, 21 respondents provided additional information revealing the content of the courses they took. As that additional information is not mandatory, few respondents answered. The majority of answers point to \textbf{Automated Testing} (6 respondents - 16.7\%) and \textbf{Fundamental Concepts} (5 respondents - 13.9\%). Those answers show that the industry should invest in the practitioners’ qualifications regarding test practices and test basic concepts. Also, the industry should invest in the practitioners’ qualification to execute specific tasks, such as \textbf{Performance Testing}, \textbf{Exploratory Tests}, \textbf{Black-box Testing}, \textbf{Unit Testing}, \textbf{Functional Testing}, \textbf{Manual Testing}, \textbf{Test Case Development}, \textbf{Test Management} (for each one, 1 answer - 2.8\%), \textbf{Behaviour-Driven Development}, and \textbf{Test-Driven Development} (for each one, 2 answers - 5.6\%).

\vspace{1.0cm}
\doublebox{
\begin{minipage}[c]{23em}
\textit{\textbf{Finding 11.}}
The industry usually invests in the qualification of test practitioners, especially when they have little or no knowledge in the area. The industry also provides learning sources and experienced practitionerns to guide the new ones.
\end{minipage}} 
\vspace{0.2cm}

\subsection{Novice software testers' knowledge (RQ4)}

We asked the respondents' perception about the possibility of novice software testers performing testing activities. From 53 respondents (77.9\%) that consider novice software testers as not prepared for the job, they cited three main reasons: \textbf{(i) Insufficient Content} (29 respondents - 54.7\%), \textbf{(ii) It is not taught in academia} (18 respondents - 34.0\%), and \textbf{(iii) Lack of Practice} (15 respondents - 28.3\%). Regarding the former and the latter, the respondent \#R7 stated: \textit{``The student does not have the dimension of the importance of testing topics, and they know even less how to develop tests''}. Respondent \#R6 also stated: \textit{``Before I started working, I did not even imagine a software testing area; I imagined that the developer was responsible for testing the software.''}

Thirteen respondents (19.1\%) considered the novice software testers as prepared. For them, the universities' starting point is enough to guide the practitioners at the beginning of their careers.  Next, they should continue learning about additional content on their own to improve their knowledge. Respondent \#R44 stated that ``The theory learned at the university can be used as an initial step while acquiring knowledge about the industry's process''. However, the respondents emphasized that extra training is necessary to support their learning process, as respondent \#R30 mentioned: \textit{``We are all prepared for the market, as long as they offer adequate training. We leave the academia knowing only generalities on software testing [...]''}. 

Two respondents (2.9\%) considered that the novice software testers could or not be either prepared, depending on the tasks. A novice software tester may perform a manual test with fewer difficulties than an automated test. For example, respondent \#R9 has affirmed, \textit{``It depends on the type of test that the tester has to create. For manual tests, I think the content presented in the university was sufficient. For other types of tests, no. The testers must learn by themselves to execute their tasks.''}

\vspace{0.4cm}
\doublebox{
\begin{minipage}[c]{23em}
\textit{\textbf{Finding 12.}}
Despite the software testing topics academia addresses guide novice software testers, they need industry-backed support to qualify themselves.
\end{minipage}} 
\vspace{0.4cm}

\vspace{0.2cm}
\doublebox{
\begin{minipage}[c]{23em}
\textit{\textbf{Finding 13.}}
The novice software testers may not be ready for entrance into the software testing industry. 
\end{minipage}}
\vspace{0.2cm}
\section{Implications for practice}
\label{sec:Implications}

Our findings suggest gaps between the topics studied in academia and practices applied in the industry. The suggestions to help to fill in this gap include the following.

\vspace{0.2cm}
\noindent
\textbf{Recognizing the software testing importance in the students' formation (Findings 1, 2, 3, 5, 7, 8, and 13)}. {\color{black}The Brazilian universities could provide practical exercises with tool support covering the core concepts in software testing. The curriculum could aggregate software testing as a specific discipline to deal with the lack of time for test practices in computing-related courses \cite{Melo2020}.}

\vspace{0.2cm}
\noindent
\textbf{In-depth study of subjects related to tests (Findings 9 and 10)}. The testing concepts, methodologies, approaches, and techniques should focus on the testing teaching process. Since the industry is constantly evolving, fundamental knowledge is essential to prepare the students for upcoming approaches. The focus should not be on specific testing tools as the use of those tools depends on the projects' context. The testing teaching process should introduce different examples for the students, presenting the concepts that underlie the tools, the benefits, and how to use them.

\vspace{0.15cm}
\noindent
\textbf{Software testing practice (Findings 6 and 10)}. The testing lessons should include examples from the industry to help the students have a real experience. Using those examples in practice activities is essential to help the students understand the theory and know-how to deal with unexpected situations. The lessons could have enriching discussions because the students would study test code examples and different standards. Furthermore, those examples could improve the ability to assimilate information and understand the test code logic - a soft skill expected by the industry.

\vspace{0.15cm}
\noindent
\textbf{Industry participation in training students (Findings 11 and 12)}. Emerging the students in the industry environment for performing software testing practices can collaborate in their formation due to the industry processes knowledge sharing. On the other hand, the industry could have programs with courses to encourage undergraduate students to be involved with their software testing code and processes, allowing the experienced practitioners to mentor novice students.

\vspace{0.15cm}
\noindent
\textbf{Integrated perspective of software testing within programming classes (Findings 4, 5, and 10)}. The practitioners need to understand the system behavior and semantics under test for developing good test cases. Therefore, the integration of software testing and programming classes can help the students develop skills for software analysis and comprehension.

\vspace{0.15cm}
\noindent
\textbf{Insertion of professors in an industrial context for training (Findings 4, 5, and 10)}. In general, the software testing professors have a high level of academic research knowledge \cite{Melo2020}. The professors may also lack practical examples (representing real projects) to bring the theoretical content closer to the industry practices. Therefore, the professors' collaboration in industrial projects may facilitate the relationship between theory and practice to improve the software testing teaching-learning process.

\section{Related Work}
\label{sec:RelatedWork}

Several studies investigate the gap between industry and academia from different points of view. Some studies investigated: i) the practitioners' perspective on industry-academia collaborations \cite{GarousiFelderer2017}, ii) the educators' perspective on software testing teaching \cite{Melo2020}, and iii) the graduates' curriculum-based perspective on software testing. There is no study focusing on the practitioners' perspective about their software testing learning in academia to the best of our knowledge.   

\citet{GarousiFelderer2017} surveyed 105 practitioners from five countries to investigate what industry expects from academia in software testing. They asked practitioners about the challenges faced through testing activities and their opinions on relevant topics to the research community. As a result, most of the respondents indicated they do not face any challenge to perform testing activities, which might imply that they may have fewer difficulties when they achieve a higher maturity. Additionally, the practitioners pose potential industry topics for research projects that converge to our survey results. For example, they highlighted the lack of training for tools usage, test automation, and lack of critical thinking as topics that academia should focus on improving the industry's practices.

\citet{Paschoal2018} surveyed 53 Brazilian professors, and \citet{Melo2020} surveyed 72 professors from 25 countries to investigate the classroom's software testing teaching. In both studies, the authors asked professors about their knowledge and experience on software testing, the concepts taught in the classroom, and the teaching approaches and support mechanisms they use. The results of both studies converge. The professors and practitioners from our survey reported similar difficulties and challenges. They have problems finding practical examples and usage of new technologies, for instance.

\citet{scatalon2018} investigated the graduates' curriculum-based perspective on software testing. They surveyed 90 Brazilian professionals regarding the testing activities they perform in their current job. They extracted a list of testing topics from a textbook \cite{del2016} and presented it to the respondents. They also calculated the gap between the topics from the textbook and the practitioners' testing activities. Similarly, our results pointed to a severe knowledge gap on the following topics: integration testing, system testing, functionality testing, performance testing, web application testing, and regression testing. Unlike our study, the results pointed to a vast knowledge in two topics: mutation testing and structural testing. This result either denotes a loss of knowledge retention on mutation testing or the current industry practices currently require more knowledge on this topic.
\section{Threats to Validity}
\label{sec:Threats}

\textbf{Internal validity.} To minimize the potential effects of selection bias, we sent the questionnaire to practitioners from different Brazilian states (response rate of 46\%). It is worth highlighting that we relied on the veracity of the respondents' information and personal opinions. Another threat that might affect internal validity refers to how we carried out the coding for open-ended questions. In this case, we conducted a peer-review data analysis to make it impartial. 

\textbf{Construct validity.} We designed the questionnaire considering a compiled of the Brazilian universities curriculum \cite{benitti2012}. However, we considered top-rated universities in Brazil that addressed the software engineering undergraduate course topics. Therefore, the software testing terminologies, types, approaches, and techniques can diverge among the universities. We mitigated that threat by providing a document with an explanation for all topics we asked the practitioners.

\textbf{External validity.} Towards mitigating the non-replicability of the study, the survey procedure and data are publicly available \cite{data}. Besides, the Brazilian universities based their curriculum on the ACM curriculum guidelines \cite{ACM2015}, which may allow the replicability of this study in other countries that follow a similar curriculum.
\section{Concluding Remarks}
\label{sec:ConcludingRemarks}

This study aimed to gather data from software testing practitioners to investigate their perception of academic learning and the industry's expectations regarding novice software testers. We conducted an expert survey with 68 practitioners who graduated from Brazilian universities. The results indicate that the novice software testers' lack of knowledge may not be related to the university maintainer, curriculum of a specific computing-related course, or course completion time. Their learning rate is similar to all scenarios. The surveyed informed to have studied 45.7\% of the software testing contents gathered through this survey. Therefore, novice software testers usually rely on industry-backed extracurricular courses to complement their learning. We synthesized the findings in the paper to make them more comprehensive and operationalizable for educators in software testing. 

As future work, we intend to conduct a global survey to understand the current gaps in academic learning and industry practices on software testing worldwide. We want to investigate how to expand the implications we propose to a practical context. In this sense, we could suggest practical examples and a practical course in software testing approaching the main topics listed in this survey. Additionally, we intend to compare our results with other studies to understand the teaching-learning process considering the different perspectives deeply: undergraduate students, test professionals, and professors.

\begin{acks}
We want to thank Ana Regina Rocha (COPPE/UFRJ), who helped us spread the questionnaire. This research was partially funded by INES 2.0; CNPq
grants 465614/2014-0 and 408356/2018-9 and FAPESB grants JCB0060/2016 and BOL0188/2020.
\end{acks}

\bibliographystyle{ACM-Reference-Format}
\bibliography{bibliography}

\end{document}